\documentclass[preprint,eqsecnum,aps,showpacs]{revtex4}
\usepackage{bm}

\begin{document}

\count255=\time\divide\count255 by 60 \xdef\hourmin{\number\count255}
  \multiply\count255 by-60\advance\count255 by\time
 \xdef\hourmin{\hourmin:\ifnum\count255<10 0\fi\the\count255}

\newcommand{\xbf}[1]{\mbox{\boldmath $ #1 $}}

\newcommand{\sixj}[6]{\mbox{$\left\{ \begin{array}{ccc} {#1} & {#2} &
{#3} \\ {#4} & {#5} & {#6} \end{array} \right\}$}}

\newcommand{\threej}[6]{\mbox{$\left( \begin{array}{ccc} {#1} & {#2} &
{#3} \\ {#4} & {#5} & {#6} \end{array} \right)$}}

\title{Baryon Magnetic Moments in Alternate $1/N_c$ Expansions}

\author{Richard F. Lebed}
\email{Richard.Lebed@asu.edu}

\author{Russell H. TerBeek}
\email{r.terbeek@asu.edu}


\affiliation{Department of Physics, Arizona State University, 
Tempe, AZ 85287-1504}

\date{November, 2010}


\begin{abstract}
Recent work shows not only the necessity of a $1/N_c$ expansion to
explain the observed mass spectrum of the lightest baryons, but also
that at least two distinct large $N_c$ expansions, in which quarks
transform under either the color fundamental or the two-index
antisymmetric representation of SU($N_c$), work comparably well.  Here
we show that the baryon magnetic moments do not support this
ambivalence; they strongly prefer the color-fundamental $1/N_c$
expansion, providing experimental evidence that nature decisively
distinguishes among $1/N_c$ expansions for this observable.
\end{abstract}

\pacs{11.15.Pg, 13.40.Em, 14.20.-c}

\maketitle
\thispagestyle{empty}

\newpage
\setcounter{page}{1}

\section{Introduction} \label{sec:Intro}

The $1/N_c$ expansion of QCD~\cite{'tHooft:1973jz} associated with the
strong interaction gauge group SU($N_c$) has in the past few decades
produced a remarkable number of insights into both the qualitative and
semi-quantitative aspects of fundamental strong interaction physics.
The existence of numerous narrow meson states, the preference of
scattering processes for channels with a minimum number of
intermediate hadronic states, the suppression of glueball-meson
mixing, the heaviness of baryons compared to mesons, the baryon mass
spectrum, and the existence of baryon resonance multiplets are
findings that now belong to the litany of standard large $N_c$
hadronic lore.

The number $N_c$ of distinct color charges [or alternately, the rank
of SU($N_c$)] is intrinsically linked to the number of valence quarks
in the baryon, which was the key observation in the original proposal
of the color quantum number~\cite{Greenberg:1964pe}.  One may
naturally suppose that quarks, being the fundamental matter
constituents of the baryons, each carry a unit of color charge in the
fundamental (F) representation of SU($N_c$), in the same way that
spin-$\frac 1 2$ fermions, being the most elementary matter particles
in the Standard Model, transform under the smallest nontrivial
representation of the Lorentz group.  However, even after imposing
this assignment at $N_c \! = \! 3$ (as is abundantly verified by the
agreement of innumerable experiments with innumerable calculations in
the asymptotic freedom regime of QCD), other generalizations are
possible for $N_c \! > \!  3$.  Note, for instance, that for $N_c \! =
\! 3$ the (anti)fundamental and two-index antisymmetric (AS)
representations of SU($N_c$) are equivalent, which is clearly seen
from the identification of $\overline{\rm F}$ and AS quark fields,
$q_i = \frac 1 2 \epsilon_{ijk} q^{j k}$.  In other words, at $N_c \!
= \! 3$ an antigreen quark is formally equivalent to a red-blue quark.
Gluons are assumed as usual to carry one color and one anticolor index
in the color adjoint representation.  For $N_c \!  > \!  3$ the
$\overline{\rm F}$ and AS color representations are distinct, meaning
that although both theories possess an SU($N_c$) gauge group, they
differ starkly in the details of their dynamics.  Each theory has a
distinct large $N_c$ limit and hence a distinct $1/N_c$ expansion
(denoted here as $1/N_c^{\rm F}$ and $1/N_c^{\rm AS}$, respectively),
as can be seen for instance by noting that quark lines carry only one
color index in the $N_c^F$ theory but two color indices in the
$N_c^{\rm AS}$ theory, and therefore internal quark loops are
suppressed by $1/N_c^1$ compared to gluon loops in $1/N_c^{\rm F}$ but
not in $1/N_c^{\rm AS}$.

While the $1/N_c^{\rm AS}$ expansion is only one of many possible
generalizations of 3-color QCD, it carries a particular theoretical
distinction.  As shown in Ref.~\cite{Armoni:2003gp}, an orientifold
equivalence relates a large class of observables between the large
$N_c^{\rm AS}$ and large $N_c^{\rm Adj}$ limits, where QCD$_{\rm Adj}$
is the corresponding generalization of QCD to a Yang-Mills theory in
which Majorana quarks carry color-adjoint representation charges.  In
turn, the theory QCD$_{\rm Adj}$ with a single massless flavor of
quark is supersymmetric, allowing the application of powerful SUSY
theorems to its analysis.  To what extent this chain of
correspondences conveys valuable phenomenological insights into the
physical case of $N_c \! = \! 3$ with several flavors of massive quark
remains an open and interesting question, but the step of analyzing
whether the $1/N_c^{\rm AS}$ expansion {\it per se\/} is supported or
denied by phenomenology can be pursued independently of these lofty
aspirations.

We investigate in this paper the viability of the $1/N_c^{\rm AS}$
expansion for a class of baryon observables, namely, their magnetic
moments.  Baryon wave functions of course depend upon the color
structure of the quarks.  Using traditional F quarks, the construction
of a color-singlet state from $N_c$ quarks is
straightforward~\cite{Witten:1979kh}:
\begin{equation} \label{Witten}
B_{\mathrm{F}} \sim \epsilon_{i_1 , \, i_2, \, \ldots \, , \, i_{N_c}}
q^{i_1} q^{i_2} \! \cdots q^{i_{N_c}} \, .
\end{equation}
The construction of color-singlet baryon states from AS quarks, on the
other hand, is neither obviously unique nor simple to express in
closed form.  Nevertheless, a construction
exists~\cite{Bolognesi:2006ws} for any value of $N_c$ that combines
$N_c(N_c \! - \! 1)/2$ AS quarks into a form fully antisymmetric under
the exchange of any two of them.  For $N_c \! = \! 3$, this expression
reads
\begin{equation}
B_{\rm AS} \sim (\epsilon_{i_2,j_2,i_1}\epsilon_{i_3,j_3,j_1}-
\epsilon_{i_3,j_3,i_1}\epsilon_{i_2,j_2,j_1})
q^{i_1 ,j_1}q^{i_2,j_2}q^{i_3 ,j_3} \, ,
\label{psi}
\end{equation}
and for all $N_c$ the corresponding expression contains two
Levi-Civita tensors.  In both the F and AS cases, each allowed quark
color combination appears precisely once in the wave function; these
numbers are ${}_{N_c} C_{1\vphantom{^\dagger}} \! = \! N_c$ for F and
${}_{N_c} C_{2\vphantom{^\dagger}} \!  = \!  N_c (N_c \! - \! 1)/2$
for AS\@.  Moreover, baryon masses scale with the number of quarks:
$\sim \! N_c$ for large $N_c^{\rm F}$ and $\sim \! N_c^2$ for large
$N_c^{\rm AS}$.  The issue of $N_c$ scaling of interactions among the
quarks in the AS case~\cite{Cherman:2006iy,Cohen:2009wm} is somewhat
more subtle than in the F case; as an example, while the exchange of a
single gluon between two F quarks introduces a factor of $g_s^2
\propto 1/N_c$~\cite{'tHooft:1973jz}, in order to maintain the color
neutrality of the full baryon state in the interaction of two AS
quarks, in a typical case a gluon must be exchanged between each of
the quarks' two color lines, leading to a factor of $g_s^4 \! \propto
\! 1/N_c^2$.

Since the quark fields comprising the fermionic baryons are completely
antisymmetrized under color in both cases, the baryon
spin-flavor-space wave functions are completely symmetric.  Under the
assumption that baryons in the ground-state multiplet have spatially
symmetric wave functions, their spin-flavor wave functions are also
completely symmetric.  Indeed, in either large $N_c$ limit, an
emergent spin-flavor
symmetry~\cite{emergent,DM,Dashen:1993jt,Carone:1993dz} collects
baryon states into spin-flavor multiplets, and the baryon observables
satisfy relations that hold to various orders in $1/N_c$.  The
ground-state multiplet for the three lightest ($u, d, s$) quark
flavors is the large $N_c$ generalization of the old SU(6) {\bf
56}-plet containing the familiar spin-$\frac 1 2$ octet and
spin-$\frac 3 2$ decuplet baryons.  The full analysis of the mass
spectrum of this multiplet in the $1/N_c^{\rm F}$ expansion appeared
many years ago in Refs.~\cite{JL}, with the result that each operator
expected to contribute to the observed baryon masses at a given order
$1/N_c^n$ does indeed produce an effect equal to $1/3^n$ times a
coefficient of order unity.  Ignoring the $N_c \! = 3$ suppressions
imposed by the $1/N_c^{\rm F}$ expansion leads to a far worse
accounting of the experimental mass spectrum.

Recently, the same techniques as in Refs.~\cite{JL} were applied to
the ground-state baryon masses using the $1/N_c^{\rm AS}$
expansion~\cite{Cherman:2009fh}.  Following the scaling and counting
arguments discussed above, one might naively expect that changing from
the $1/N_c^{\rm F}$ to the $1/N_c^{\rm AS}$ expansion would simply
induce the modification $1/N_c^n \! \to \! 1/N_c^{2n}$, rescaling
operator coefficients by powers of 3 for $N_c \! = \! 3$, and thus
spoil the agreement.  Remarkably, the $1/N_c^{\rm AS}$ fits turn out
to be of comparable quality to those for $1/N_c^{\rm F}$, and both are
far superior to fits with no $1/N_c$ suppressions at all; one
concludes that nature requires {\em some\/} $1/N_c$ expansion, but
does not---at least from the baryon mass spectrum---indicate which one
is preferred.

The purpose of the current work is to determine whether the
$1/N_c^{\rm AS}$ expansion provides a viable explanation of the
ground-state baryon magnetic moment spectrum.  After the masses, the
magnetic moments provide the largest set of precisely measured baryon
static observables.  They also provide a nearly orthogonal set of
information to the masses, since they are strongly dependent upon the
charges of the component quarks: Even baryons in a single isospin
multiplet such as the proton and neutron, which differ chiefly by the
substitution of a single quark $u \leftrightarrow d$, carry rather
different values of magnetic moment, approximately related by the
famous SU(6) result $\mu_n = -2\mu_p /3$.

Baryon magnetic moments in the $1/N_c^{\rm F}$ expansion were first
considered in Refs.~\cite{Jenkins:1994md,Luty:1994ub,Dai:1995zg}.  In
each case, a set of operators considered physically most significant
to the observables were included while others were neglected.  Further
work focused on the $\Delta \! \to \! N \gamma$
transition~\cite{Jenkins:2002rj}.  A complete basis in the $1/N_c^{\rm
F}$ expansion was not produced until much later, in
Ref.~\cite{Lebed:2004fj}.  This work explored two $1/N_c^{\rm F}$
expansion variants: one in which the sole parameter organizing the
expansion is $1/N_c$, and one a more physical {\it single-photon
ansatz}, in which the flavor structure respects that magnetic moments
involve couplings of a baryon to a single photon at lowest order, and
thus each quark should couple proportionally to its electric charge.

In this paper we work entirely within the single-photon ansatz and
compute matrix elements of a set of operators truncated at a
consistent order in both the $1/N_c^{\rm F}$ and $1/N_c^{\rm AS}$
expansions.  While Ref.~\cite{Lebed:2004fj} shows that one may compute
a complete set of operator matrix elements, a full analysis such as
performed for the masses is not currently possible since many of the
baryon magnetic moments (particularly for the decuplet and strange
decuplet-octet transitions) remain unmeasured.  As a result, we
produce a fit to coefficients at as high of an order in the $1/N_c$
expansions as possible, given current data.  As shown below, the fits
in the $1/N_c^{\rm F}$ and $1/N_c^{\rm AS}$ expansions are not both
consistent with data to comparable confidence; in particular, the
$1/N_c^{\rm AS}$ fit generates effects too large to be consistent with
the $1/N_c^{\rm AS}$ expansion, while the corresponding quantities in
the $1/N_c^{\rm F}$ expansion are all of a natural size, and using no
$1/N_c$ expansion at all would predict these quantities to be
anomalously small.  That the the spectrum of baryon magnetic moments
appears to require and prefer one particular $1/N_c$ expansion
strongly is is the conclusion of this work.

This paper is organized as follows: In Sec.~\ref{Sec:Operators} we
reprise the operator basis for the $1/N_c^{\rm F}$ expansion used in
Ref.~\cite{Lebed:2004fj}, both for the pure and single-photon ansatz
$1/N_c$ expansions.  In Sec.~\ref{Sec:NcAS} we detail the
modifications necessary to carry out the analogous analysis in the
$1/N_c^{\rm AS}$ expansion.  Results of fits to magnetic moment
observables and a discussion of their significance appear in
Sec.~\ref{Sec:Results}, and we summarize in
Sec.~\ref{Sec:Conclusions}.

\section{Operator Bases} \label{Sec:Operators}

The enumeration of independent baryon magnetic moment operators and
the calculation of their matrix elements in the $1/N_c^{\rm F}$
expansion are discussed in detail in Secs.~II and III of
Ref.~\cite{Lebed:2004fj}.  We summarize here the essential points,
inasmuch as they are germane to providing a point of comparison to the
calculation in the $1/N_c^{\rm AS}$ case to be described in the next
section.

The method by which observables classified by their spin-flavor
properties can be calculated for the ground-state multiplet of baryons
in the $1/N_c$ expansion has been understood for many years, in the
current context dating back to Ref.~\cite{Dashen:1994qi}, with similar
approaches appearing in Refs.~\cite{Carone:1993dz,Luty:1993fu}.  One
defines a complete set of spin, flavor, and spin-flavor operators that
act upon the quarks comprising the baryon using the building blocks:
\begin{eqnarray}
J^i & \equiv & \sum_\alpha q^\dagger_\alpha \left( \frac{\sigma^i}{2}
\otimes \openone \right) q_\alpha \, , \nonumber \\
T^a & \equiv & \sum_\alpha q^\dagger_\alpha \left( \openone \otimes
\frac{\lambda^a}{2} \right) q_\alpha \, , \nonumber \\
G^{ia} & \equiv & \sum_\alpha q^\dagger_\alpha \left(
\frac{\sigma^i}{2} \otimes \frac{\lambda^a}{2} \right) q_\alpha \, ,
\end{eqnarray}
where the index $\alpha$ sums over all the quarks in the baryon,
$\sigma^i$ are the Pauli spin matrices, and $\lambda^a$ are the
Gell-Mann flavor matrices.  This set of operators spans all possible
spin-flavor actions upon a single quark ({\it one-body operators}); in
a baryon containing $M$ quarks [$M \! = \! N_c$ in the $1/N_c^{\rm F}$
expansion, $M \! = \! N_c(N_c \! - \!  1)/2$ in the $1/N_c^{\rm AS}$
expansion], the most general operator linearly independent from those
appearing at lower orders requires a polynomial in one-body operators
no higher than degree $M$ ({\it i.e.}, $0 \! \le \! n \! \le \!
M$-body operators).  For the magnetic moments of the ground-state
multiplet baryons, one forms all independent operators transforming as
T odd, $\Delta J \! = \! 1$, $\Delta J_3 \!  = \! 0$, $\Delta Y \! =
\! 0$, and $\Delta I_3 \! = \!  0$~\footnote{Magnetic moment photonic
couplings do not necessarily transform as $\Delta J_3 \! = \! 0$, but
all couplings with $\Delta J_3 \! \neq \! 0$ are precisely related to
them by the exact SU(2) rotational symmetry.}; their number must
precisely equal that of the distinct observables carrying these
quantum numbers.

Many of the operators produced in this fashion are linearly dependent,
particularly when acting upon the completely symmetric ground-state
multiplet (For example, any operator acting antisymmetrically on two
quarks annihilates such states).  The full {\em operator reduction
rule\/} for removing all such superfluous operators acting upon the
ground-state multiplet appears in Ref.~\cite{Dashen:1994qi}, and has
already been applied to the set of operators appearing here.

Once the basis is established, one must take into account both
explicit and implicit factors of $N_c$.  The explicit powers arise as
overall scaling due to the 't~Hooft power counting; an $n$-body
operator requires at minimum the exchange of $(n \! - \! 1)$ [$2(n \!
- \! 1)$] gluons between the $n$ quark lines in the $1/N_c^{\rm F}$
($1/N_c^{\rm AS}$) expansion, leading to a scaling coefficient of
$1/N_c^n$ for the $1/N_c^{\rm F}$ expansion, $1/N_c^{2n}$ for the
$1/N_c^{\rm AS}$ expansion.  Implicit factors of $N_c$ arise due to
combinatorics; contributions from the $M$ quarks can add coherently
for many of the operators to give factors of $M^n$ in their matrix
elements.

Finally, since operator matrix elements can contain both leading and
subleading contributions in $N_c$, it is possible for a set of
operators to be linearly independent when all terms are included but
linearly dependent when only the leading terms are retained.  As an
example, the operators $\openone$ and $T^8$ both have $O(M^1)$ matrix
elements, but their linear combination $N_s \! \equiv \! \frac 1 3 (
\openone \! - \! 2 \sqrt{3} T^8 )$ simply counts the number of strange
quarks in a baryon state, and hence has matrix elements of $O(M^0)$
when applied to the familiar ground-state baryon states.  Such
suppressed linear combinations are called {\it demoted
operators\/}~\cite{Carlson:1998gw}.  A proper enumeration of operators
in either $1/N_c$ expansion includes the effect of all demotions.

Even a list of operators satisfying all of these conditions is
unnecessary; since one ultimately compares the results of the
calculation to $N_c \! = \! 3$ baryon states, only an expansion up to
and including 3-body operators is required for a full accounting of
data.  A complete set of operators in the $1/N_c^{\rm F}$ expansion up
to 3-body level, after accounting for all scaling and combinatoric
powers of $N_c$, and imposing the restrictions of operator reduction
rules and demotions, appears as Table~I of Ref.~\cite{Lebed:2004fj},
and is reproduced for convenience as Table~\ref{oplist} here.  One
notes that the list contains precisely 27 operators, which matches the
number of baryon magnetic moment observables in the ground-state
multiplet of spin-$\frac 1 2$ and spin-$\frac 3 2$ baryons: A magnetic
moment for each of the octet and decuplet baryons, and transitions
between the 9 pairs of states with the same values of electric charge
and strangeness ({\it i.e.}, $\Sigma^0 \Lambda$, $\Delta^+ p$, {\it
etc.}).  The full set of matrix elements for these 27 operators
evaluated for the 27 observables occupies Tables~IV--IX in
Ref.~\cite{Lebed:2004fj}.
\begin{table}
\caption{From Ref.~\cite{Lebed:2004fj}: The 27 linearly independent
operators contributing to the magnetic moments of the spin-$\frac 1 2$
and spin-$\frac 3 2$ ground-state baryons, organized according to the
leading $N_c$ counting of their matrix elements in the $1/N_c^{\rm F}$
expansion.\label{oplist}}
\begin{tabular}{r|l}
\hline\hline
$O(N_c^1)$ \ & \ $G^{33}$ \\
$O(N_c^0)$ \ & \ $J^3$, \ $G^{38}$, \ $\frac{1}{N_c} T^3 G^{33}$, \
$\frac{1}{N_c} N_s G^{33}$, \ $\frac{1}{N_c^2} \frac 1 2 \{ J^i G^{i3},
G^{33} \}$ \\
$O(N_c^{-1})$ \ & \ $\frac{1}{N_c} T^3 J^3$, \ $\frac{1}{N_c} N_s J^3$,
\ $\frac{1}{N_c} T^3 G^{38}$, \ $\frac{1}{N_c} N_s G^{38}$, \
$\frac{1}{N_c^2} \frac 1 2 \{ J^2, G^{33} \}$, \ $\frac{1}{N_c^2}
(T^3)^2 G^{33}$, $\frac{1}{N_c^2} N_s^2 G^{33}$, \\
& \ $\frac{1}{N_c^2} T^3 N_s G^{33}$, \ $\frac{1}{N_c^2} J^i G^{i3}
J^3$, \ $\frac{1}{N_c^2} \frac 1 2 \{ J^i G^{i8}, G^{33} \}$, \
$\frac{1}{N_c^2} \frac 1 2 \{ J^i G^{i3}, G^{38} \}$ \\
$O(N_c^{-2})$ \ & \ $\frac{1}{N_c^2} J^2 J^3$, \
$\frac{1}{N_c^2} N_s^2 J^3$, \ $\frac{1}{N_c^2} (T^3)^2 J^3$, \
$\frac{1}{N_c^2} T^3 N_s J^3$, \
$\frac{1}{N_c^2} \frac 1 2 \{ J^2, G^{38} \}$, \
$\frac{1}{N_c^2} (T^3)^2 G^{38}$, $\frac{1}{N_c^2} N_s^2 G^{38}$, \\
& \ $\frac{1}{N_c^2} T^3 N_s G^{38}$, \
$\frac{1}{N_c^2} J^i G^{i8} J^3$, \
$\frac{1}{N_c^2} \frac 1 2 \{ J^i G^{i8}, G^{38} \}$ \\
\hline
\end{tabular}
\end{table}

However, as indicated above we adopt a different organization of the
operator basis, founded on the physical assumption (the single-photon
ansatz) that the quarks in any magnetic moment operator couple
proportionally to their electric charges.  That is, one assumes the
magnetic moment operator couples to a single photon through the flavor
combinations
\begin{eqnarray}
Q = T^Q & \equiv & T^3 + \frac{1}{\sqrt{3}} T^8 , \nonumber \\
G^{iQ} & \equiv & G^{i3} + \frac{1}{\sqrt{3}} \, G^{i8} \, .
\end{eqnarray}
Under this assumption, the only operators appearing up to 3-body level
(obtained as linear combinations of those in Table~\ref{oplist}) with
no other source of SU(3) flavor breaking are given
by~\cite{Lebed:2004fj}
\begin{equation} \label{LO}
{\cal O}_1 \! \equiv G^{3Q}, \ {\cal O}_2 \equiv \frac{1}{N_c} Q
J^3, \ \tilde {\cal O}_3 \equiv \! \frac{1}{N_c^2} \frac 1 2 \{
J^2, G^{3Q} \} , \ {\cal O}_4 \equiv \! \frac{1}{N_c^2} J^i G^{iQ}
J^3 \, ,
\end{equation}
which give matrix elements of $O(N_c^1)$, $O(N_c^0)$, $O(N_c^{-1})$,
and $O(N_c^{-1})$, respectively.  The operator combination
\begin{equation} \label{O3def}
{\cal O}_3 \equiv \! (\tilde {\cal O}_3 \! - {\cal O}_4)
\end{equation}
has the interesting property that it vanishes for all diagonal
magnetic moments and therefore provides particularly incisive
information about the transition moments.  The list of leading-order
operators in the $1/N_c^{\rm F}$ single-photon ansatz is therefore
given by ${\cal O}_{1,2,3,4}$.

Additional sources of SU(3) flavor breaking include negligibly small
effects due to the presence of a second (loop) photon (proportional to
$\alpha/{4\pi}$) or to the difference $(m_u \!  - \! m_d)$.  However,
the dominant additional SU(3)-flavor breaking effects occur due to the
distinction of the strange quark, $m_s \! \gg \! m_{u,d}$, and are
indicated by the presence of an SU(3)-breaking parameter $\varepsilon$
expected to be $\sim \! 0.3$.  Such phenomena are manifested either
through the strangeness-counting operator $N_s$ or the strange quark
spin operator,
\begin{equation}
J_s^i \equiv \frac 1 3 (J^i \! - 2\sqrt{3} G^{i8}) \, .
\end{equation}
Even for the operators in this category, the couplings are still
assumed to follow the single-photon ansatz and therefore either
include one $Q$ or $G^{3Q}$ operator, or one power of the strange
quark charge $q_s$.  At $O(\varepsilon^1 N_c^0)$, one then obtains the
additional operators:
\begin{equation} \label{NLO}
\varepsilon {\cal O}_5 \equiv \varepsilon q_s J_s^3 , \
\varepsilon {\cal O}_6 \equiv \! \frac{\varepsilon}{N_c} N_s G^{3Q} , \
\varepsilon {\cal O}_7 \equiv \! \frac{\varepsilon}{N_c} \, Q J_s^3 \, ,
\end{equation}
and at $O(\varepsilon^1 N_c^{-1})$, one finds the operators:
\begin{eqnarray}
&&
\varepsilon {\cal O}_8 \equiv \varepsilon q_s \frac{N_s}{N_c} J^3 , \
\varepsilon {\cal O}_9 \equiv \varepsilon \frac{N_s}{N_c^2} Q J^3 , \
\varepsilon {\cal O}_{10} \equiv \frac{\varepsilon}{N_c^2} \frac 1 2
\{ {\bf J} \! \cdot \! {\bf J}_s , G^{3Q} \}, \
\nonumber \\ &&
\varepsilon {\cal O}_{11} \equiv \frac{\varepsilon}{N_c^2} J_s^j
G^{jQ} J^3, \
\varepsilon {\cal O}_{12} \equiv \frac{\varepsilon}{N_c^2} \frac 1 2
\{ J^j G^{jQ}, J_s^3 \} \, . \label{NNLO}
\end{eqnarray}
Beyond this point, the next operators would have matrix elements of
$O(\varepsilon^2 N_c^{-1})$ or $O(\varepsilon^1 N_c^{-2})$.  However,
none appear at $O(\varepsilon^0 N_c^{-2})$ because the list of
operators in Eq.~(\ref{LO}) is exhaustive to the 3-body level, and
none appear at $O(\varepsilon^2 N_c^0)$ because $\varepsilon^2$
implies at least a 2-body operator, whose matrix elements on the
physical baryon states are at most $O(1/N_c^1)$.  These observations
have important consequences for choosing consistent truncation points
in the combined expansion in $\varepsilon$ and $1/N_c$, which we
discuss in detail in the next section.

The operators ${\cal O}_{1,2, \ldots, 12}$ were defined in
Ref.~\cite{Lebed:2004fj}.  While their matrix elements evaluated on
the 27 observables for $N_c \! = \! 3$ form a rank-12 matrix (and
hence are independent), they were not explicitly tabulated in
Ref.~\cite{Lebed:2004fj}, and it was not noticed at the time that two
combinations of them are demotable.  To wit, $\frac 1 2 \varepsilon
{\cal O}_8 + \varepsilon {\cal O}_9$ has matrix elements of
$O(\varepsilon^1 N_c^2)$ and hence should be neglected in this
analysis ({\it i.e.}, $\varepsilon {\cal O}_9$ can be eliminated in
favor of $\varepsilon {\cal O}_8$ at this order), and the combination
\begin{equation} \label{O13def}
\varepsilon {\cal O}_{13} \equiv \frac 1 2 \varepsilon {\cal O}_5 +
\varepsilon {\cal O}_7
\end{equation}
has matrix elements of $O(\varepsilon^1 N_c^{-1})$ while each of
$\varepsilon {\cal O}_5$ and $\varepsilon {\cal O}_7$ have matrix
elements of $O(\varepsilon^1 N_c^0)$, and hence $\varepsilon {\cal
O}_{13}$ belongs to the same list as Eq.~(\ref{NLO}), while
$\varepsilon {\cal O}_7$ can be eliminated in favor of $\varepsilon
{\cal O}_5$.  The full set of matrix elements for all 27 observables
for all of ${\cal O}_{1, \ldots, 13}$ in the $1/N_c^{\rm F}$ expansion
are presented in Tables~\ref{S2a}--\ref{S2c}.

\begin{table}
\caption{Matrix elements of magnetic moment operators defined in
Eqs.~(\ref{LO})--(\ref{O3def}).\label{S2a}}
\begin{tabular}{c|c|c|c|c}
\hline\hline
State & $\langle {\cal O}_1 \rangle$ & $\langle {\cal O}_2 \rangle$ &
$\langle {\cal O}_3 \rangle$ & $\langle {\cal O}_4 \rangle$ \\
\hline\hline
$\Delta^{++}$ & $\frac{1}{20} (3N_c + 11)$ & $\frac{1}{4N_c}(N_c + 9)$
& $0$ & $\frac{3}{16N_c^2} (3N_c + 11)$ \\
$\Delta^+$ & $\frac{1}{20} (N_c + 7)$ & $\frac{1}{4N_c}(N_c + 3)$ &
$0$ & $\frac{3}{16N_c^2} (N_c + 7)$ \\
$\Delta^0$ & $-\frac{1}{20} (N_c - 3)$ & $\frac{1}{4N_c}(N_c - 3)$ &
$0$ & $-\frac{3}{16N_c^2} (N_c - 3)$ \\
$\Delta^-$ & $-\frac{1}{20} (3N_c + 1)$ & $\frac{1}{4N_c}(N_c - 9)$ &
$0$ & $-\frac{3}{16N_c^2} (3N_c + 1)$ \\
$\Sigma^{*+}$ & $\frac{1}{8} (N_c + 1)$ & $\frac{1}{4N_c}(N_c + 3)$ &
0 & $\frac{15}{32N_c^2} (N_c + 1)$ \\
$\Sigma^{*0}$ & $0$ & $\frac{1}{4N_c} (N_c - 3)$ & $0$ & $0$ \\
$\Sigma^{*-}$ & $-\frac{1}{8} (N_c + 1)$ & $\frac{1}{4N_c} (N_c - 9)$
& $0$ & $-\frac{15}{32N_c^2} (N_c + 1)$ \\
$\Xi^{*0}$ & $\frac{1}{12} (N_c - 3)$ & $\frac{1}{4N_c} (N_c - 3)$ &
$0$ & $\frac{5}{16N_c^2} (N_c - 3)$ \\
$\Xi^{*-}$ & $-\frac{1}{12} (N_c + 3)$ & $\frac{1}{4N_c} (N_c - 9)$ &
$0$ & $-\frac{5}{16N_c^2} (N_c + 3)$ \\
$\Omega^-$ & $-\frac{1}{2}$ & $\frac{1}{4N_c} (N_c - 9)$ & $0$ &
$-\frac{15}{8N_c^2}$ \\
$p$ & $\frac{1}{12} (N_c + 3)$ & $\frac{1}{12N_c} (N_c + 3)$ & $0$ &
$\frac{1}{16N_c^2} (N_c + 3)$ \\
$n$ & $-\frac{1}{12} (N_c + 1)$ & $\frac{1}{12N_c} (N_c - 3)$ & $0$ &
$-\frac{1}{16N_c^2} (N_c + 1)$ \\
$\Sigma^+$ & $\frac{1}{12} (N_c + 3)$ & $\frac{1}{12N_c} (N_c + 3)$ &
$0$ & $\frac{1}{16N_c^2} (N_c + 3)$ \\
$\Sigma^0$ & $\frac{1}{6}$ & $\frac{1}{12N_c} (N_c - 3)$ & $0$ &
$\frac{1}{8N_c^2}$ \\
$\Lambda$ & $-\frac{1}{6}$ & $\frac{1}{12N_c} (N_c - 3)$ & $0$ &
$-\frac{1}{8N_c^2}$ \\
$\Sigma^0 \Lambda$ & $-\frac{1}{12} \! \sqrt{(N_c - 1)(N_c + 3)}$ &
$0$ & $0$ & $-\frac{1}{16N_c^2}\sqrt{(N_c - 1)(N_c + 3)}$ \\
$\Sigma^-$ & $-\frac{1}{12} (N_c - 1)$ & $\frac{1}{12N_c} (N_c - 9)$ &
$0$ & $-\frac{1}{16N_c^2} (N_c - 1)$ \\
$\Xi^0$ & $-\frac{1}{36} (N_c + 9)$ & $\frac{1}{12N_c} (N_c - 3)$ &
$0$ & $-\frac{1}{48N_c^2} (N_c + 9)$ \\
$\Xi^-$ & $\frac{1}{36} (N_c - 9)$ & $\frac{1}{12N_c} (N_c - 9)$ & $0$
& $\frac{1}{48N_c^2} (N_c - 9)$ \\
$\Delta^+ p$ & $\frac{1}{6\sqrt{2}} \sqrt{(N_c - 1)(N_c + 5)}$ & $0$ &
$\frac{3}{8\sqrt{2}N_c^2} \sqrt{(N_c - 1)(N_c + 5)}$ & $0$ \\
$\Delta^0 n$ & $\frac{1}{6\sqrt{2}} \sqrt{(N_c - 1)(N_c + 5)}$ & $0$ &
$\frac{3}{8\sqrt{2}N_c^2} \sqrt{(N_c - 1)(N_c + 5)}$ & $0$ \\
$\Sigma^{*0} \Lambda$ & $\frac{1}{6\sqrt{2}} \sqrt{(N_c - 1) (N_c +
3)}$ & $0$ & $\frac{3}{8\sqrt{2}N_c^2}\sqrt{(N_c - 1)(N_c + 3)}$ & $0$
\\
$\Sigma^{*0} \Sigma^0$ & $\frac{1}{3\sqrt{2}}$ & $0$ &
$\frac{3}{4\sqrt{2}N_c^2}$ & $0$ \\
$\Sigma^{*+} \Sigma^+$ & $\frac{1}{12\sqrt{2}}(N_c + 5)$ & $0$ &
$\frac{3}{16\sqrt{2}N_c^2}(N_c + 5)$ & $0$ \\
$\Sigma^{*-} \Sigma^-$ & $-\frac{1}{12\sqrt{2}}(N_c - 3)$ & $0$ &
$-\frac{3}{16\sqrt{2}N_c^2} (N_c - 3)$ & $0$ \\
$\Xi^{*0} \Xi^0$ & $\frac{1}{9\sqrt{2}}(N_c + 3)$ & $0$ &
$\frac{1}{4\sqrt{2}N_c^2}(N_c + 3)$ & $0$ \\
$\Xi^{*-} \Xi^-$ & $-\frac{1}{9\sqrt{2}}(N_c - 3)$ & $0$ &
$-\frac{1}{4\sqrt{2}N_c^2}(N_c - 3)$ & $0$ \\
\hline
\end{tabular}
\end{table}

\begin{table}
\caption{Matrix elements of magnetic moment operators defined in
Eqs.~(\ref{NLO})--(\ref{NNLO}).\label{S2b}}
\begin{tabular}{c|c|c|c|c|c}
\hline\hline
State & $\langle {\cal O}_5 \rangle$ & $\langle {\cal O}_6 \rangle$ &
$\langle {\cal O}_7 \rangle$ & $\langle {\cal O}_8 \rangle$ & $\langle
{\cal O}_9 \rangle$ \\
\hline\hline
$\Delta^{++}$ & $0$ & $0$ & $0$ & $0$ & $0$ \\
$\Delta^+$ & $0$ & $0$ & $0$ & $0$ & $0$ \\
$\Delta^0$ & $0$ & $0$ & $0$ & $0$ & $0$ \\
$\Delta^-$ & $0$ & $0$ & $0$ & $0$ & $0$ \\
$\Sigma^{*+}$ & $-\frac{1}{6}$ & $\frac{1}{8N_c} (N_c + 1)$ &
$\frac{1}{12N_c} (N_c + 3)$ & $-\frac{1}{2N_c}$ & $\frac{1}{4N_c^2}
(N_c + 3)$ \\
$\Sigma^{*0}$ & $-\frac{1}{6}$ & $0$ & $\frac{1}{12N_c} (N_c - 3)$ &
$-\frac{1}{2N_c}$ & $\frac{1}{4N_c^2} (N_c - 3)$ \\
$\Sigma^{*-}$ & $-\frac{1}{6}$ & $-\frac{1}{8N_c} (N_c + 1)$ &
$\frac{1}{12N_c} (N_c - 9)$ & $-\frac{1}{2N_c}$ & $\frac{1}{4N_c^2}
(N_c - 9)$ \\
$\Xi^{*0}$ & $-\frac{1}{3}$ & $\frac{1}{6N_c} (N_c - 3)$ &
$\frac{1}{6N_c} (N_c - 3)$ & $-\frac{1}{N_c}$ & $\frac{1}{2N_c^2} (N_c
- 3)$ \\
$\Xi^{*-}$ & $-\frac{1}{3}$ & $-\frac{1}{6N_c} (N_c + 3)$ &
$\frac{1}{6N_c} (N_c - 9)$ & $-\frac{1}{N_c}$ & $\frac{1}{2N_c^2} (N_c
- 9)$ \\
$\Omega^-$ & $-\frac{1}{2}$ & $-\frac{3}{2N_c}$ & $\frac{1}{4N_c} (N_c
- 9)$ & $-\frac{3}{2N_c}$ & $\frac{3}{4N_c^2} (N_c - 9)$ \\
$p$ & $0$ & $0$ & $0$ & $0$ & $0$ \\
$n$ & $0$ & $0$ & $0$ & $0$ & $0$ \\
$\Sigma^+$ & $\frac{1}{18}$ & $\frac{1}{12N_c} (N_c + 3)$ &
$-\frac{1}{36N_c} (N_c + 3)$ & $-\frac{1}{6N_c}$ & $\frac{1}{12N_c^2}
(N_c + 3)$ \\
$\Sigma^0$ & $\frac{1}{18}$ & $\frac{1}{6N_c}$ & $-\frac{1}{36N_c}
(N_c - 3)$ & $-\frac{1}{6N_c}$ & $\frac{1}{12N_c^2} (N_c - 3)$ \\
$\Lambda$ & $-\frac{1}{6}$ & $-\frac{1}{6N_c}$ & $\frac{1}{12N_c} (N_c
- 3)$ & $-\frac{1}{6N_c}$ & $\frac{1}{12N_c^2} (N_c - 3)$ \\
$\Sigma^0 \Lambda$ & $0$ & $-\frac{1}{12N_c}\sqrt{(N_c - 1)(N_c + 3)}$
& $0$ & $0$ & $0$ \\
$\Sigma^-$ & $\frac{1}{18}$ & $-\frac{1}{12N_c} (N_c - 1)$ &
$-\frac{1}{36N_c} (N_c - 9)$ & $-\frac{1}{6N_c}$ & $\frac{1}{12N_c^2}
(N_c - 9)$ \\
$\Xi^0$ & $-\frac{2}{9}$ & $-\frac{1}{18N_c} (N_c + 9)$ &
$\frac{1}{9N_c} (N_c - 3)$ & $-\frac{1}{3N_c}$ & $\frac{1}{6N_c^2}
(N_c - 3)$ \\
$\Xi^-$ & $-\frac{2}{9}$ & $\frac{1}{18N_c} (N_c - 9)$ &
$\frac{1}{9N_c} (N_c - 9)$ & $-\frac{1}{3N_c}$ & $\frac{1}{6N_c^2}
(N_c - 9)$ \\
$\Delta^+ p$ & $0$ & $0$ & $0$ & $0$ & $0$ \\
$\Delta^0 n$ & $0$ & $0$ & $0$ & $0$ & $0$ \\
$\Sigma^{*0} \Lambda$ & $0$ & $\frac{1}{6\sqrt{2}N_c} \sqrt{(N_c -
1)(N_c + 3)}$ & $0$ & $0$ & $0$ \\
$\Sigma^{*0} \Sigma^0$ & $\frac{\sqrt{2}}{9}$ &
$\frac{1}{3\sqrt{2}N_c}$ & $-\frac{1}{9\sqrt{2}N_c} (N_c - 3)$ & $0$ &
$0$ \\
$\Sigma^{*+} \Sigma^+$ & $\frac{\sqrt{2}}{9}$ &
$\frac{1}{12\sqrt{2}N_c}(N_c + 5)$ & $-\frac{1}{9\sqrt{2}N_c}(N_c +
3)$ & $0$ & $0$ \\
$\Sigma^{*-} \Sigma^-$ & $\frac{\sqrt{2}}{9}$ &
$-\frac{1}{12\sqrt{2}N_c}(N_c - 3)$ & $-\frac{1}{9\sqrt{2}N_c}(N_c -
9)$ & $0$ & $0$ \\
$\Xi^{*0} \Xi^0$ & $\frac{\sqrt{2}}{9}$ & $\frac{\sqrt{2}}{9N_c}(N_c +
3)$ & $-\frac{1}{9\sqrt{2}N_c} (N_c - 3)$ & $0$ & $0$ \\
$\Xi^{*-} \Xi^-$ & $\frac{\sqrt{2}}{9}$ & $-\frac{\sqrt{2}}{9N_c}(N_c
- 3)$ & $-\frac{1}{9\sqrt{2}N_c} (N_c - 9)$ & $0$ & $0$ \\
\hline
\end{tabular}
\end{table}

\begin{table}
\caption{Matrix elements of magnetic moment operators defined in
Eqs.~(\ref{NNLO})--(\ref{O13def}).\label{S2c}}
\begin{tabular}{c|c|c|c|c}
\hline\hline
State & $\langle {\cal O}_{10} \rangle$ & $\langle {\cal O}_{11}
\rangle$ & $\langle {\cal O}_{12} \rangle$ & $\langle {\cal O}_{13}
\rangle$ \\
\hline\hline
$\Delta^{++}$ & $0$ & $0$ & $0$ & $0$ \\
$\Delta^+$ & $0$ & $0$ & $0$ & $0$ \\
$\Delta^0$ & $0$ & $0$ & $0$ & $0$ \\
$\Delta^-$ & $0$ & $0$ & $0$ & $0$ \\
$\Sigma^{*+}$ & $\frac{5}{32N_c^2} (N_c + 1)$ & $\frac{1}{32N_c^2}
(3N_c - 5)$ & $\frac{5}{32N_c^2} (N_c + 1)$ & $\frac{1}{4N_c}$ \\
$\Sigma^{*0}$ & $0$ & $-\frac{1}{4N_c^2}$ & $0$ & $-\frac{1}{4N_c}$ \\
$\Sigma^{*-}$ & $-\frac{5}{32N_c^2} (N_c + 1)$ & $-\frac{1}{32N_c^2}
(3N_c + 11)$ & $-\frac{5}{32N_c^2} (N_c + 1)$ & $-\frac{3}{4N_c}$
\\
$\Xi^{*0}$ & $\frac{5}{24N_c^2} (N_c - 3)$ & $\frac{1}{8N_c^2} (N_c -
7)$ & $\frac{5}{24N_c^2} (N_c - 3)$ & $-\frac{1}{2N_c}$ \\
$\Xi^{*-}$ & $-\frac{5}{24N_c^2} (N_c + 3)$ & $-\frac{1}{8N_c^2} (N_c
+ 7)$ & $-\frac{5}{24N_c^2} (N_c + 3)$ & $-\frac{3}{2N_c}$ \\
$\Omega^-$ & $-\frac{15}{8N_c^2}$ & $-\frac{15}{8N_c^2}$ &
$-\frac{15}{8N_c^2}$ & $-\frac{9}{4N_c}$ \\
$p$ & $0$ & $0$ & $0$ & $0$ \\
$n$ & $0$ & $0$ & $0$ & $0$ \\
$\Sigma^+$ & $-\frac{1}{48N_c^2} (N_c + 3)$ & $-\frac{1}{48N_c^2}
(3N_c + 13)$ & $-\frac{1}{48N_c^2} (N_c + 3)$ & $-\frac{1}{12N_c}$
\\
$\Sigma^0$ & $-\frac{1}{24N_c^2}$ & $-\frac{5}{24N_c^2}$ &
$-\frac{1}{24N_c^2}$ & $\frac{1}{12N_c}$ \\
$\Lambda$ & $-\frac{1}{8N_c^2}$ & $-\frac{1}{8N_c^2}$ & $-\frac{1}{8N_c^2}$ &
$-\frac{1}{4N_c}$ \\
$\Sigma^0 \Lambda$ & $-\frac{1}{48N_c^2} \sqrt{(N_c - 1)(N_c + 3)}$ &
$-\frac{1}{16N_c^2} \sqrt{(N_c - 1)(N_c + 3)}$ & $-\frac{1}{48N_c^2}
\sqrt{(N_c - 1)(N_c + 3)}$ & $0$ \\
$\Sigma^-$ & $\frac{1}{48N_c^2} (N_c - 1)$ & $\frac{1}{48N_c^2} (3N_c
- 7)$ & $\frac{1}{48N_c^2} (N_c - 1)$ & $\frac{1}{4N_c}$ \\
$\Xi^0$ & $-\frac{1}{36N_c^2} (N_c + 9)$ & $-\frac{1}{12N_c^2} (N_c +
5)$ & $-\frac{1}{36N_c^2} (N_c + 9)$ & $-\frac{1}{3N_c}$  \\
$\Xi^-$ & $\frac{1}{36N_c^2} (N_c - 9)$ & $\frac{1}{12N_c^2} (N_c -
5)$ & $\frac{1}{36N_c^2} (N_c - 9)$ & $-\frac{1}{N_c}$ \\
$\Delta^+ p$ & $0$ & $0$ & $0$ & $0$ \\
$\Delta^0 n$ & $0$ & $0$ & $0$ & $0$ \\
$\Sigma^{*0} \Lambda$ & $\frac{1}{6\sqrt{2}N_c^2} \sqrt{(N_c - 1)(N_c
+ 3)}$ & $0$ & $\frac{1}{24\sqrt{2}N_c^2} \sqrt{(N_c - 1)(N_c + 3)}$ &
$0$ \\
$\Sigma^{*0} \Sigma^0$ & $\frac{1}{6\sqrt{2}N_c^2}$ & $0$ &
$-\frac{1}{12\sqrt{2}N_c^2}$ & $\frac{1}{3\sqrt{2}N_c}$ \\
$\Sigma^{*+} \Sigma^+$ & $\frac{1}{24\sqrt{2}N_c^2} (N_c + 5)$ & $0$ &
$-\frac{1}{48\sqrt{2}N_c^2} (7N_c + 11)$ & $-\frac{1}{3\sqrt{2}N_c}$
\\
$\Sigma^{*-} \Sigma^-$ & $-\frac{1}{24\sqrt{2}N_c^2} (N_c - 3)$ & $0$
& $\frac{1}{48\sqrt{2}N_c^2} (7N_c + 3)$ & $\frac{1}{\sqrt{2}N_c}$ \\
$\Xi^{*0} \Xi^0$ & $\frac{7}{36\sqrt{2}N_c^2} (N_c + 3)$ & $0$ &
$-\frac{1}{18\sqrt{2}N_c^2} (N_c - 6)$ & $\frac{1}{3\sqrt{2}N_c}$ \\
$\Xi^{*-} \Xi^-$ & $-\frac{7}{36\sqrt{2}N_c^2} (N_c - 3)$ & $0$ &
$\frac{1}{18\sqrt{2}N_c^2} (N_c + 6)$ & $\frac{1}{\sqrt{2}N_c}$ \\
\hline
\end{tabular}
\end{table}

\section{The $1/N_c^{\rm AS}$ Expansion} \label{Sec:NcAS}

Calculating the corresponding matrix elements in the $1/N_c^{\rm AS}$
expansion is remarkably straightforward once one possesses their
values in the $1/N_c$ expansion.  Since baryons are fermions and thus
have wave functions completely antisymmetric under the exchange of any
two quarks, and the baryon wave functions constructed from both F and
AS quarks are completely antisymmetrized under the exchange of any two
quarks in color space, the spin-flavor-space wave functions are
completely symmetric.  In the ground-state multiplet, which by
assumption is completely symmetric in spatial coordinates, the
spin-flavor wave functions must also be completely symmetric.
Precisely the same symmetrization condition holds for baryons built
from either F or AS quarks; the operators in spin-flavor space carry
precisely all the same indices in either case, and all of the results
on the definition and interpretation of various operators, what
operator reduction rules they obey, which combinations are demoted,
{\it etc.}, carry over {\it mutatis mutandis}.

In fact, only two changes need to be made in order to apply the
results of the previous section and the results of
Tables~\ref{S2a}--\ref{S2c} to the $1/N_c^{\rm AS}$ expansion.  As
mentioned there, the powers of $N_c$ due to 't~Hooft scaling are
changed from $1/N_c^n$ in the $1/N_c^{\rm F}$ expansion to
$1/N_c^{2n}$ in the $1/N_c^{\rm AS}$ expansion, and the combinatoric
factors due to the number of quarks (called $M$ in the previous
section) are changed from $N_c^1$ in the $1/N_c^{\rm F}$ expansion to
$N_c (N_c \! - \! 1)/2$ in the $1/N_c^{\rm AS}$ expansion.  Since the
former and the latter factors are clearly segregated in
Tables~\ref{S2a}--\ref{S2c}, a simple substitution generalizes their
application from QCD$_{\rm F}$ to QCD$_{\rm AS}$.

One may now address the question of consistent truncation of the
operator basis in powers of $\varepsilon$ and $1/N_c$ in the two
expansions.  In order to proceed, one must determine the relative
parametric size of $\varepsilon$ compared to $1/N_c$.  Traditionally,
$\varepsilon$ is estimated numerically from the relative size of
strangeness mass splittings in a baryon multiplet or the size of
departures of strange hadron couplings from their SU(3) symmetric
values, or in chiral perturbation theory as effects of
$O(m_s/\Lambda_{\chi})$, where $\Lambda_{\chi}$ is the chiral
symmetry-breaking scale.  Such effects are estimated to be no more
than about 30\%, {\it i.e.}, parametrically equal to $1/N_c$.  As
discussed in Ref.~\cite{Lebed:2004fj}, however, chiral perturbation
theory also contains SU(3)-violating loop corrections of
$O(m_s^{1/2})$, suggesting that, in some cases, $\varepsilon \! \sim
\! 1/N_c^{1/2}$.  Reference~\cite{Lebed:2004fj} then showed that using
either scaling of $\varepsilon$, the operator expansion for the
$1/N_c^{\rm F}$ expansion may be consistently truncated either
including ${\cal O}_{1, \ldots , 7}$ [including effects up through
$O(\varepsilon^1 N_c^0)$ and $O(\varepsilon^0 N_c^{-1})$] or including
${\cal O}_{1, \ldots , 13}$ [including up through $O(\varepsilon^1
N_c^{-1})$ while neglecting $O(\varepsilon^2 N_c^{-1})$ and
$O(\varepsilon^1 N_c^{-2})$].  Remarkably, for either scaling of
$\varepsilon$ the $1/N_c^{\rm AS}$ expansion may be truncated after
the same sets of operators: The set ${\cal O}_{1, \ldots , 7}$
includes effects up through $O(\varepsilon^1 N_c^0)$ and
$O(\varepsilon^0 N_c^{-2})$, while the set ${\cal O}_{1, \ldots , 13}$
includes effects up through $O(\varepsilon^1 N_c^{-2})$ while
neglecting $O(\varepsilon^2 N_c^{-2})$ and $O(\varepsilon^1
N_c^{-4})$.

\section{Results of Fits to Measured Moments} \label{Sec:Results}

While the ground-state baryon multiplet contains 27 magnetic moment
observables, many of them have never been measured chiefly due to the
fact that most of the decuplet states are strongly decaying
resonances, for which detecting electromagnetic processes is extremely
difficult.  9 are tabulated in the {\em Review of Particle
Physics}~\cite{Nakamura:2010zzi}: magnetic moments of 7 of the 8 octet
baryons ($\mu_{\Sigma^0}$ is unknown), the $\Omega^-$, and the
$\Sigma^0 \Lambda$ transition moment.  The $\Delta^{\! +} p$
transition moment can be extracted from the $\Delta \! \to \! N
\gamma$ helicity amplitudes and is found to be $\mu_{\Delta^{\! +} p} \!
= \! 3.51 \pm 0.09 \, \mu_N$.  We also use the extracted value
$\mu_{\Delta^{\! ++}} \! = 6.14 \pm 0.51 \,
\mu_N$~\cite{LopezCastro:2000ep}, which is not the sole value used in
Ref.~\cite{Nakamura:2010zzi} and is obtained from an analysis of data
that has some model dependence, but the extraction performed respects
both gauge invariance and the finite $\Delta^{\! ++}$ width.  The
tabulated $\Delta^+$ moment value~\cite{Kotulla:2002cg},
$\mu_{\Delta^{\! +}} \! = 2.7^{+ 1.0}_{-1.3} \: {\rm (stat)} \pm 1.5
\: {\rm (syst)} \pm 3 \: {\rm (theor)} \, \mu_N$ has such a large
theoretical uncertainty that we do not use it in our fits.

We have seen in Sec.~\ref{Sec:NcAS} that the full set of operators up
to and including $O(\varepsilon^1 N_c^{-1})$ in the $1/N_c^{\rm F}$
expansion, or $O(\varepsilon^1 N_c^{-2})$ in the $1/N_c^{\rm AS}$
expansion, consists only of the 11 operators ${\cal
O}_{1,2,3,4,5,6,8,10,11,12,13}$ defined in Sec.~\ref{Sec:Operators}.
In fact, when restricted to the 11 observed moments, two more
combinations among these operators are demoted, as may be verified
through a quick check of Table~\ref{Sec:Operators}: Each of
$\varepsilon {\cal O}_{10,11,12}$ has matrix elements of
$O(\varepsilon^1 N_c^{-1})$ in the $1/N_c^{\rm F}$ expansion
[$O(\varepsilon^1 N_c^{-2})$ in the $1/N_c^{\rm AS}$ expansion], but
the combinations $(-\frac 1 3 {\cal O}_{11} \! + {\cal O}_{12})$ and
$({\cal O}_{10} \! - {\cal O}_{12})$ have matrix elements of
$O(\varepsilon^1 N_c^{-2})$ [$1/N_c^{\rm F}$] or $O(\varepsilon^1
N_c^{-4})$ [$1/N_c^{\rm AS}$], which should be neglected in our
consistent-order expansion.  Thus, ${\cal O}_{11}$ and ${\cal O}_{12}$
may be eliminated in favor of ${\cal O}_{10}$, leaving a basis of only
9 operators, ${\cal O}_{1,2,3,4,5,6,8,10,13}$.

One may ask whether it is appropriate to include terms both leading
and subleading in $N_c$ in the matrix elements.  We argue that such
terms are essential to reproduce the complete physical nature of
electromagnetic interactions such as magnetic moments.  Consider, for
example, the operator $Q$\@.  Since the physical ground-state baryons
contain $M$ quarks [either $N_c$ or $N_c(N_c \! - \! 1)/2$] but differ
in no more than the 3 valence quarks, the leading $O(M^1)$
contribution from $Q$ is the same for all of the physical ground-state
baryons; {\it i.e.}, all of the observed baryons have the same
electric charge at leading order in $N_c$.  This peculiar result
arises from the independence of quark electric charges $q_u \! = \!
+\frac 2 3$, $q_d \! = \! q_s \! = \! -\frac 1 3$ (a manifestly
electromagnetic effect) from $N_c$ scaling (a strong interaction
effect).  Since we insist that our results extrapolate in a physically
meaningful way from $N_c \! = \! 3$, we retain subleading terms in the
matrix elements for our fits.

The 11 observed moments are fit to a set of 9 $O(N_c^0)$ operator
coefficients $d_{i_n} \! = \! d_{1,2,3,4,5,6,8,10,13}$ used to define
the full magnetic moment operator:
\begin{equation} \label{newexp}
\mu_z = \mu_0 \sum_{n=1}^9 d_{i_n} \,
\varepsilon^{k_{i_n}} {\cal O}_{i_n} \, ,
\end{equation}
where, as indicated by Eqs.~(\ref{LO})--(\ref{O13def}), $k_{i_n} \! =
\!  0$ for $i_n \! = \! 1, \ldots , 4$ and $k_{i_n} \! = \! 1$ for
$i_n \! = \! 5, \ldots , 13$.  As in Ref.~\cite{Lebed:2004fj}, we set
the overall scale $\mu_0$ to equal $2\mu_p$ in order to make the
coefficient $d_1$ of the sole leading-order operator, ${\cal O}_1$, of
order unity for $N_c \! = \! 3$.  Since Eq.~(\ref{newexp}) neglects
all $O(\varepsilon^2/M)$ and $O(\varepsilon/M^2)$ contributions to the
magnetic moments, one must combine the statistical uncertainty of each
moment with a ``theoretical uncertainty'' of magnitude given by the
larger of $O(\mu_p \, \varepsilon^2/M)$ and $O(\mu_p \,
\varepsilon/M^2)$.

Using $N_c \! = \! 3$ and $\varepsilon \! = \! \frac 1 3$, one obtains
the coefficients for the $1/N_c^{\rm F}$ expansion in
Table~\ref{NcFfit} and those for the $1/N_c^{\rm AS}$ expansion in
Table~\ref{NcASfit}.  The $\chi^2$/d.o.f.\ for the $1/N_c^{\rm F}$
expansion is 0.31, and that for the $1/N_c^{\rm AS}$ is 1.55.  If the
scale of the theoretical uncertainty uses $2\mu_p$ rather than
$\mu_p$, these numbers drop to 0.09 and 0.61, respectively, meaning
that the quality of the fit is good in either case, and therefore the
values obtained for the operator coefficients are reliably determined.
A glance at Table~\ref{NcFfit} shows that every coefficient in the
$1/N_c^{\rm F}$ neatly assumes a value of $O(1)$ or less, thus
following the dictates of either $1/N_c$ expansion.  Furthermore,
unlike the fit to only 7 operators in Ref.~\cite{Lebed:2004fj} in
which most of the coefficients are anomalously suppressed, only $d_2$
appears to have an especially small coefficient.  On the other hand,
the coefficients $d_3$, $d_4$, $d_8$, and $d_{10}$ in the $1/N_c^{\rm
AS}$ fit are substantially larger than $O(1)$, indicating the failure
of the $1/N_c^{\rm AS}$ expansion for the magnetic moments.

One can also check that the obvious candidates for rescuing the
$1/N_c^{\rm AS}$ expansion are inadequate to the task.  Using, as
discussed above, either the $1/N_c^{\rm F}$ or $1/N_c^{\rm AS}$
expansion without including subleading terms leads to the prediction
of some coefficients of unnaturally large size, while using the
alternate parameter choice $\varepsilon \! \sim \!  N_c^{-1/2}$ rather
than $\varepsilon \! \sim \!  N_c^{-1}$ has relatively little effect
on the pattern of coefficient magnitudes.  Meanwhile, neglecting the
$1/N_c$ expansion entirely---achieved by deleting all scaling powers
of $1/N_c$ from Eqs.~(\ref{LO})--(\ref{O13def}) while setting
combinatoric factors of $N_c$ to 3---leads to a fit given in
Table~\ref{NcNonefit} [here, $\chi^2/{\rm d.o.f.}$ is only 0.039
because the theoretical uncertainty is now $O(\mu_p \,
\varepsilon^2)$] with several coefficients $d_{i_k}$ substantially less
than $O(1)$ ($d_2$, $d_3$, $d_4$, and likely others when uncertainties
are taken into account): A $1/N_c$ expansion is clearly needed to
explain their natural sizes.  The $1/N_c^F$ expansion in the
single-photon ansatz including subleading terms appears to have a
``Goldilocks'' quality: While the other $1/N_c$ expansions produce
coefficients too large and ignoring the $1/N_c$ expansion entirely
produces coefficients too small, the $1/N_c^F$ expansion thus far is
unique in producing uniformly ``just right'' $O(1)$ coefficients.
\begin{table}
\caption{Best fit values for the coefficients in the $1/N_c^{\rm F}$
expansion using Eq.~(\ref{newexp}).\label{NcFfit}}
\begin{tabular}{rrr}
\hline\hline
$d_1    = +0.992 \pm 0.044$ \  & $d_2    = -0.078 \pm 0.148$ \ &
$d_3    = +1.363 \pm 0.272$ \\
$d_4    = +0.461 \pm 0.489$ \  & $d_5    = -1.652 \pm 0.566$ \ &
$d_6    = -0.288 \pm 0.438$ \\
$d_8    = +1.588 \pm 0.865$ \  & $d_{10} = -3.727 \pm 2.852$ \ &
$d_{13} = +0.499 \pm 0.438$ \\
\hline
\end{tabular}
\end{table}
\begin{table}
\caption{Best fit values for the coefficients in the $1/N_c^{\rm AS}$
expansion using Eq.~(\ref{newexp}).\label{NcASfit}}
\begin{tabular}{rrr}
\hline\hline
$d_1    = +0.976  \pm 0.023$ \  & $d_2    = -0.188 \pm 0.176$ \ &
$d_3    = +12.846 \pm 1.553$ \\
$d_4    = +5.289  \pm 2.743$ \  & $d_5    = -1.474 \pm 0.223$ \ &
$d_6    = -1.147  \pm 0.491$ \\
$d_8    = +4.841  \pm 1.046$ \  & $d_{10} = -36.332 \pm 12.322$ \ &
$d_{13} = +1.218  \pm 0.490$ \\
\hline
\end{tabular}
\end{table}
\begin{table}
\caption{Best fit values for the coefficients using no $1/N_c$
expansion in Eq.~(\ref{newexp}).\label{NcNonefit}}
\begin{tabular}{rrr}
\hline\hline
$d_1    = +0.995 \pm 0.116$ \  & $d_2    = -0.029 \pm 0.138$ \ &
$d_3    = +0.150 \pm 0.075$ \\
$d_4    = +0.051 \pm 0.121$ \  & $d_5    = -1.708 \pm 1.593$ \ &
$d_6    = -0.085 \pm 0.420$ \\
$d_8    = +0.535 \pm 0.829$ \  & $d_{10} = -0.420 \pm 0.845$ \ &
$d_{13} = +0.178 \pm 0.420$ \\
\hline
\end{tabular}
\end{table}

The fit values for the coefficients given in Table~\ref{NcFfit} may be
used to predict all 16 unknown magnetic moment observables, as done in
Ref.~\cite{Lebed:2004fj}; the results are compiled in
Table~\ref{pred}~\footnote{In particular, the predicted value
$\mu_{\Delta^+}$ is easily compatible with the given experimental
value.}.  The difference is that the operator basis has been fit here
including effects of $O(\varepsilon^1 N_c^{-1})$, but only including
$O(\varepsilon^1 N_c^0)$ and $O(\varepsilon^0 N_c^{-1})$ in
Ref.~\cite{Lebed:2004fj}, and so the theoretical uncertainty for the
best-determined moments here is $O(\mu_p \, \varepsilon^1 N_c^{-2})$
or $O(\mu_p \, \varepsilon^2 N_c^{-1})$.  For the strange decuplet
moments or strange decuplet-octet transitions, however, at least one
of the combinations $(-\frac 1 3 {\cal O}_{11} \! + \! {\cal O}_{12})$
and $({\cal O}_{10} \!  - \! {\cal O}_{12})$ is no longer demoted, so
that the fit values for those moments must include the larger
theoretical uncertainty of $O(\mu_p \, \varepsilon^1 N_c^{-1})$.  The
results presented in Table~\ref{pred} improve upon, and in almost all
cases agree within 1$\sigma$ with, the results in Table~XI of
Ref.~\cite{Lebed:2004fj}.
\begin{table}
\caption{Best fit values for the 16 unknown magnetic moments in units
of $\mu_N$ using the $1/N_c^{\rm F}$ expansion.\label{pred}}
\begin{tabular}{rrrr}
\hline\hline
$\mu_{\Delta^{\! +}}    = +3.09 \pm 0.16$ & \
$\mu_{\Delta^{\! 0}}    = +0.00 \pm 0.10$ & \
$\mu_{\Delta^{\! -}}    = -3.09 \pm 0.16$ & \
$\mu_{\Sigma^{*+}}      = +2.62 \pm 0.35$ \\
$\mu_{\Sigma^{*0}}      = -0.06 \pm 0.32$ & \
$\mu_{\Sigma^{*-}}      = -2.73 \pm 0.35$ & \
$\mu_{\Xi^{*0}}         = -0.12 \pm 0.33$ & \
$\mu_{\Xi^{*-}}         = -2.37 \pm 0.39$ \\
$\mu_{\Sigma^0}         = +0.65 \pm 0.11$ & \
$\mu_{\Delta^{\! 0} n}  = +3.51 \pm 0.11$ & \
$\mu_{\Sigma^{*0} \Lambda}  = +2.65 \pm 0.32$ & \
$\mu_{\Sigma^{*0} \Sigma^0} = +1.21 \pm 0.31$ \\
$\mu_{\Sigma^{*+} \Sigma^+} = +2.69 \pm 0.32$ & \
$\mu_{\Sigma^{*-} \Sigma^-} = -0.26 \pm 0.31$ & \
$\mu_{\Xi^{*0} \Xi^0}       = +2.30 \pm 0.33$ & \
$\mu_{\Xi^{*-} \Xi^-}       = -0.26 \pm 0.31$ \\
\hline
\end{tabular}
\end{table}

Since we fit 9 operator coefficients using 11 observables, one can
also investigate the two magnetic moment combinations satisfied by all
the operators.  One of them has been known since the early days of
heavy baryon chiral perturbation theory~\cite{Jenkins:1992pi}:
\begin{equation} \label{reln1}
\mu_n -\frac 1 4 (\mu_{\Sigma^+} + \mu_{\Sigma^-}) -\frac 3 2
\mu_{\Lambda} -\sqrt{3} \mu_{\Sigma^0 \Lambda} + \mu_{\Xi^0} =
O(\mu_p \, \varepsilon^2 M^{-1}) , O(\mu_p \, \varepsilon^1 M^{-2}) \,
.
\end{equation}
Experimentally, this combination is $(0.22 \pm 0.14) \mu_N$.  The best
scale-independent measure for this relation is obtained by dividing it
by the average of the same combination with all negative values
replaced by positive ones; such a combination is $O(\mu_p \, M^1)$,
and their ratio is therefore predicted to be the larger of
$O(\varepsilon^2 M^{-2})$ and $O(\varepsilon^1 M^{-3})$.  One obtains
$0.057 \pm 0.036$ {\it vs.}\ 0.012 ($1/N_c^{\rm F}$) or 0.0014
($1/N_c^{\rm AS}$).  While it is tempting to ascribe a superior
agreement to the $1/N_c^{\rm F}$ expansion, in fact the central
experimental value is only $1.6\sigma$ from zero (due almost entirely
to the large $\mu_{\Sigma^0 \Lambda}$ uncertainty), so that neither
expansion is particularly favored for this single result.  This
conclusion is even more stark for the other relation:
\begin{equation} \label{reln2}
\mu_{\Delta^{++}} + 2 \mu_{\Omega^-} + 8 ( \mu_p + \mu_n ) - 6
\mu_{\Sigma^+} + 12 \mu_\Lambda + 4 \mu_{\Sigma^-} - 12 \mu_{\Xi^0} -
2 \mu_{\Xi^-} = O(\mu_p \, \varepsilon^2 M^{-1}) , O(\mu_p \,
\varepsilon^1 M^{-2}) \, ,
\end{equation}
for which the experimental value $(-1.30 \pm 0.56) \mu_N$ converts to
the ratio $-0.029 \pm 0.012$.  In this case, the central value is
about 2.3$\sigma$ from zero, and the $1/N_c^{\rm F}$ expansion is
somewhat preferred; reducing the large $\mu_{\Delta^{++}}$ uncertainty
would sharpen this conclusion.

\section{Conclusions} \label{Sec:Conclusions}
The remarkable fact that the baryon mass spectrum not only requires a
$1/N_c$ expansion in order to explain the size of its suppressed
combinations but works about as well for more than one such possible
expansion---$1/N_c^{\rm F}$ and $1/N_c^{\rm AS}$---does not survive
the scrutiny of results from baryon magnetic moments.  We have shown
with fits to the observed data that the $1/N_c^{\rm F}$ expansion
produces no unnaturally large coefficients in the $1/N_c$ expansion of
the magnetic moment operator, while such large coefficients are
unavoidable in the $1/N_c^{\rm AS}$ expansion.  Moreover, numerous
coefficients in a fit that entirely ignores the $1/N_c$ expansion are
highly suppressed, mandating a $1/N_c$ expansion in order to satisfy
the naturalness criterion.

Such a result might seem surprising; after all, the baryon mass data
set is more complete than that of the magnetic moments.  However, as
remarked above, the mass spectrum and magnetic moments probe largely
orthogonal physical effects.  The sole leading-order [$O(M^1)$, with
$M \! \sim \! N_c^1$ for $1/N_c^{\rm F}$, $N_c^2$ for $1/N_c^{\rm
AS}$] operator $\openone$ for the mass spectrum simply gives the same
universal mass to all baryons, while the sole leading-order operator
$G^{3Q}$ for the magnetic moments gives contributions that vary from
state to state.  Thus it is perhaps not so surprising that a more
incisive result arises from the magnetic moment sector.

In addition, this work advances the analysis of the baryon magnetic
moments to one order higher in the $1/N_c$ expansion [now including
all $O(\varepsilon^1 N_c^{-1})$ effects] than was previously known.
Issues about numerous potentially small coefficients have largely
evaporated in light of the new analysis.  Should the measured but less
well-known moments, particularly $\mu_{\Sigma^0 \Lambda}$,
$\mu_{\Delta^{++}}$, and $\mu_{\Delta^+}$, receive renewed
experimental scrutiny, and should more radiative decays of strange
decuplet baryons be observed~\cite{Taylor:2005zw} and
analyzed~\cite{Lebed:2004zc}, the $1/N_c$ expansion will be subject to
ever more precise tests of its applicability.

\section*{Acknowledgments}
This work was supported in part by the National Science Foundation
under Grant No.\ PHY-0757394.

\end{document}